\documentclass[12pt]{iopart}

\begin{document}
\maketitle
\title[Third Order Integrable Equation Possessing Symplectic Operator Of Degree 9]{Third Order Integrable Equation Possessing Symplectic Operator Of Degree 9}

\author[Daryoush Talati]{Daryoush Talati}

\ead{talati@eng.ankara.edu.tr , daryoush.talati@gmail.com}

\begin{abstract}
We perform a classification of third order integrable systems of evolution equations with respect to higher symmetries. Applying it, we consider polynomial systems that are 0-homogeneous under a suitable weighting of variables with main matrix $\left(\begin{array}{cc} 3\sqrt{5}+5 &0 \\ 0&-3\sqrt{5}+5\end{array}\right) $ having a seventh-order symmetry. A new integrable equation is discovered. For this new supersymmetric equation the Lax representation and bi-Hamiltonian structures are given. 

\end{abstract}

\section{Introduction}

Nonlinear equations which are solvable by a transformation to a linear equation or by an inverse spectral transformation possesses various surprising features such as infinitely many symmetries and conserved covariants, and they are integrable \cite{cal}. Such models arise in many branches of physics such as classical and quantum field theories, particle physics, relativity, statistical physics and quantum gravity. The difficulty in constructing an inverse spectral transformation had motivated the search for other methods which would identify the equations expected to be solvable by an inverse spectral transformation. These methods which consist of finding a property shared by all known integrable equations are called “integrability tests”. The existence of an infinite number of conserved quantities, infinite number of symmetries, soliton solutions, Hamiltonian and bi-Hamiltonian structures, Lax pairs, Painleve property, are well known integrability tests.

In most of integrable evolution equations, the right-hand side of the equation is a homogeneous polynomial under a suitable weighting scheme. The system of differential equation is said to be $\lambda$-homogeneous of weight $\tau$ if it admits the one-parameter group of scaling symmetries
\begin{eqnarray}\fl
(x,t,u,v)\longrightarrow(a^{-1}x, a^{-\tau}t, a^{\lambda}u, a^{\lambda}v)~,~~~ a \in \Re^{+}~~.
\end {eqnarray}
In this work we are restricting our attention to $0$-homogeneous equations.

Let us consider a system of two t-independent evolution equations

\begin{eqnarray} \fl
\left(
\begin{array}{l}
u\\
v
\end{array}\right)_{t}=
\left(
\begin{array}{c} 
F_1[u,v]\\
F_2[u,v]
\label{nn1}
\end{array}
\right) 
\end{eqnarray}

where $F [u, v] = F (u, v, u_x , v_x, u_{xx}, v_{xx}, . . . )$ denotes a differential function of $u$ and $v$. A second system of evolution equations
\begin{eqnarray} \fl
\left(
\begin{array}{l}
u\\
v
\end{array}\right)_{t}=
\left(
\begin{array}{c} 
Q_1[u,v]\\
Q_2[u,v]
\label{nn}
\end{array}
\right) 
\end{eqnarray}

is said to be a generalized symmetry of (\ref{nn}) if the corresponding flows commute

\begin{eqnarray}
D_F(Q)-D_Q(F)=0.
\end {eqnarray}

Here $F=(F_1, F_2), Q=(Q_1,Q_2)$, and $D$ denotes the Frechet derivative. An equation is called integrable if it has infinitely many higher order symmetries. By the triangular system we understand such system which involves either an equation depending only on $u$ or an equation depending only on $v$. A subclass of symmetry integrable equations which are called completely integrable, further possess infinitely many conservation laws. Complete integrability of an equation is established by
proving a compatible bi-Hamiltonian, (or bi-symplectic or Hamiltonian-symplectic operators) defining the Magri scheme of infinite symmetry hierarchy $u_{t_{i}}=F_{i}[u]$ all of which is in conservation law form
\begin{eqnarray*}\fl
\begin{array}{l}
u_{t_{i}}=F_{i}[u]={\rm K}G_{i}[u]={\rm J}G_{i+1}[u], \;\; G_{i}=\mathrm{\delta}(\rho_{i}[u]),\;\;\;i=-1,0,1,2,3,\cdots
\end{array}
\end{eqnarray*}
with respect to two compatible Hamiltonian (skew-adjoint, Jacobi identity satisfying) operators $\mathrm{J}$ and $\mathrm{K}$ by the conserved density $\rho_i[u]$ which are in involution. Hamiltonian operators are defined to be compatible if their arbitrary linear combinations are Hamiltonian operators also. The dual objects $G_{i}[u]$ are the conserved gradients satisfying
\begin{eqnarray*}
{\rm D}_{F_i}(G_i)-{\rm D^{*}}_{G_i}(F_i)=0
\end {eqnarray*}
and the Helmholz condition of having self adjoint Frechet derivative 
${\rm D}_{G_i}^{*}={\rm D}_{G_i}$ which are therefore Euler derivative 
\begin{eqnarray*}\fl
\mathrm{\delta}=\sum_{n=0} (-D)^{n}\frac{\partial\phantom{Q} }{\partial u_n}
\end {eqnarray*}
of a conserved density $\rho_{i}[u]$.

If a nonlinear equation possesses a Lax pair, then the Lax pair may be used to gather information about the behavior of the solutions to the nonlinear equation. Importantly, the existence of a Lax pair is a signature of integrability of the associated nonlinear equation.The term ‘Lax pair’ refers to linear systems that are related to nonlinear equations through a compatibility condition. The first part of Lax pair is called the scattering problem, that allows the initial-value problem for the integrable equation to be solved exactly.

 In his seminal work \cite{lax}, Lax suggested a formalism to integrate a class of nonlinear evolution equations. He introduced a pair of linear operators $L$ and $M$ such that
 
\begin{eqnarray} \fl
 L\phi=\lambda \phi,~~~\phi_t=M\phi,\label{123}
 \end {eqnarray}

where L and M are linear differential operators, $\lambda$ is an eigenvalue of $L$, and $\phi$ is an eigenfunction of $L$. Assuming $\lambda_t = 0$, differentiating $L\phi$ with respect to t gives

\begin{eqnarray*} \fl
L_t\phi+L\phi_t=\lambda\phi_t.
\end {eqnarray*}

Substituting in from (\ref{123}) gives that

\begin{eqnarray*} \fl
L_t\phi=ML\phi-LM\phi.
\end {eqnarray*}
And therefore
\begin{eqnarray*} \fl
(L_t+[L,M])\phi=0.
\end {eqnarray*}
Where $[M,L]= ML - LM$ is the operator commutator. Hence,

\begin{eqnarray}\fl\begin{array}{ll}
L_t=[M,L], \label{lax}
\end{array}\end {eqnarray}

is called Lax equation and contains commutative nonlinear evolution equation for suitable $L$
and $M$. In \cite{777}, it is shown that all scalar evolution equations, which have the form of a conservation law, have a Lax pair with a second order L. For example consider the Lax formalism for the KdV equation

\begin{eqnarray}\fl\begin{array}{ll}
L=-\mathcal{D}_x^2+u+ u_1\mathcal{D}_x^{-1} ,~~~~
M=-4\mathcal{D}_x^3+6 u\mathcal{D}_x+9u_1+3 u_2\mathcal{D}_x^{-1}.\label{kdv2} 
\end{array}\end {eqnarray}

These operators satisfies the Lax equation (\ref{lax})

\begin{eqnarray*}\fl\begin{array}{ll}
L_t=u_t+ u_{tx}D_x^{-1} ,~~~~
[M,L]= u_3-6uu_1+(u_4-6uu_2-6u_1^2)D_x^{-1},
\end{array}\end {eqnarray*}

if $u$ is solution to the KdV equation:

\begin{eqnarray*}\fl
u_t= -u_3+6uu_1.
\end {eqnarray*}

Many approaches have developed to the classification of integrable equations, but the most successful one is based on the generalized symmetries \cite{a26,a3,a4,a5,a7}. Many integrable systems have now been completely classified using this direct computational approach. The scalar equations admit various generalizations to the multi field case and have been often considered in the literature \cite{f6,a16,f2000,dtt,dt}. Many years ago a system of two-component coupled equations with Jordan canonical main matrix $\left(\begin{array}{cc} 3\sqrt{5}+5 &0 \\ 0&-3\sqrt{5}+5\end{array}\right) $discovered by Drinfel’d and Sokolov \cite{f6} and later rediscovered by Sakovich and Foursov \cite{f7,f8}

\begin{eqnarray} \fl
\left(
\begin{array}{l}
u\\\\
v\\\\
\end{array}\right)_{t}=
\left(
\begin{array}{cc} 
(3\sqrt{5}+5)u_{3x}+10u_{x}u+(3\sqrt{5}+5)vu_{x}\\+2\sqrt{5}uv_{x}+(\sqrt{5}-1)vv_{x}\\
(-3\sqrt{5}+5)v_{3x}+(\sqrt{5}+5)5uu_{x}-10vu_{x}\\+(\sqrt{5}-3)5uv_{x}-2\sqrt{5}vv_{x}
\end{array}
\right) .\label{ds} 
\end{eqnarray}
The sysstem (\ref{ds}) is homogeneous if we assign weightings of $2$ to the dependent variables $u$ and $v$, while $x-$ and $t-$ differentiation have weights $1$ and $3$, respectively. 
Foursov showed that it has a remarkably unusual behavior. indeed, it does not possess symmetries of orders (at least) 5 and 15. It thus belongs to the first known symmetry hierarchy that starts with the following sequence of order 3, 7, 9, 11, 13, 17. Recently bi-Hamiltonian formulation for this equation constructed in \cite{tt}.

\section{The Main Results}

By a linear change of dependent variables a two-component system of evolution equations can be reduced to a system with the main matrix in Jordan canonical form

\begin{eqnarray*} \fl
J_1=\left(
\begin{array}{cc}
1 & 0 \\
0 & 1
\end{array}
\right),\qquad
J_2=
\left(
\begin{array}{cc}
1 & 0 \\
0 & \lambda
\end{array}
\right),\qquad
J_3=
\left(
\begin{array}{cc}
1 & \lambda \\
0 & 1
\end{array}
\right)
\end {eqnarray*}

with nonzero $\lambda$. On the other hand if there exist a integrable system of two component equations of weight 2 in the specific Jordan form, then there is at least one integrable system of same Jordan form of weight 0. Starting from sysstem (\ref{ds}), This is the motivation of the present work about the problem of classification of 0-Homogeneous evolution equations in $\left(\begin{array}{cc} 3\sqrt{5}+5 &0 \\ 0&-3\sqrt{5}+5\end{array}\right) $ canonical Jordan form for integrability.

Consider the systems of form 

\begin{eqnarray} \fl
\left(
\begin{array}{l}
u\\\\
v\\\\
\end{array}\right)_{t}=
\left(
\begin{array}{c} 

(3\sqrt{5}+5) u_{3x}+ l_1u_{x} u_{xx} + l_2u_{x} v_{xx} + l_3v_{x}u_{xx} + l_4v_{x} v_{xx}\\ + l_5 u_{x}^3 + l_6 v_{x}^3 + l_7u_{x}^2 v_{x} + l_8u_{x} v_{x}^2\\
(-3\sqrt{5}+5) v_{3x} + m_1u_{x} u_{xx} + m_2u_{x} v_{xx} + m_3v_{x}u_{xx} + m_4v_{x} v_{xx} \\ + m_5 u_{x}^3+ m_6 v_{x}^3 + m_7u_{x}^2 v_{x} + m_8u_{x} v_{x}^2

\label{nn2}
\end{array}
\right) .
\end{eqnarray}

Note that system (\ref{nn2}) is homogeneous if we assign weightings of $0$ to the dependent variables $u$ and $v$, while $x-$ and $t-$ differentiation have weights $1$ and $3$, respectively.$\\$

\emph{{\bf Theorem 2.1.}}  Any nonlinear non triangular system (\ref{nn2}), having a symmetry of order seven, up to scalings of $t, x, u, v$ can be transformed to the following system:

\begin{eqnarray} \fl
\left(
\begin{array}{l}
u\\ \\
v \\\\
\end{array}\right)_{t}=
\left(
\begin{array}{c} 
48 u_{3x}+36 v_{3x}-6 uv_{xx} -12v v_{xx} -6 u_{x} v_{x}-2u^{2} u_{x} +v^{2}u_{x} \\-12 v_{x}^{2}+2u v v_{x} \\
36 u_{3x}+12 v_{3x}+6u u_{xx} +12 vu_{xx} +6 u_{x}^{2}+12 u_{x} v_{x}+2u v u_{x} \\+v_{x} u^{2}-2 v_{x} v^{2} 
\label{nm}
\end{array}
\right) .
\end{eqnarray}

This new system has a remarkably unusual behavior too. The only known nonlinear equation admit nontrivial operator of degree $>7$. indeed,  it does not possess symmetries of orders (at least) 5 and 15. It belongs to the first known symmetry hierarchy that starts with the following sequence of order 3, 7, 9, 11, 13, 17... and also doesn't admit any reduction to any of KdV, mKdV, pKdV or Burgers equations.$\\$

\emph{{\bf Proposition 2.2.}}  The infinite hierarchy of System (\ref{nm}) have infinite conserved densities  together suffice to write two Magri \cite{mag} schemes  : 
\begin{eqnarray} \fl
\left(
\begin{array}{cc} 
u_{t_i}\\
v_{t_i}
\end{array}
\right)
=
\mathrm{J} 
\left(
\begin{array}{cc} 
\delta_{u}\\
\delta_{v}
\end{array}
\right)
\int \rho_{i-1}~ \mathrm{d}x
=\mathrm{J}\circ\mathrm{K}\circ\mathrm{J} 
\left(
\begin{array}{cc} 
\delta_{u}\\
\delta_{v}
\end{array}
\right)
\int \rho_{i-11} \mathrm{d}x\label{hm}
\end{eqnarray}
with the compatible pair of Hamiltonian operators
\begin{eqnarray} \fl
\mathrm{J} = \left(
\begin{array}{cc} 
D_x & 0 \\
 0 & D_x 
\end{array}\right), \mathrm{K} = \left(
\begin{array}{cc} 
\mathrm{K} _{1}&\mathrm{K} _{2}\\
-\mathrm{K _{2}}^{*}&\mathrm{K} _{4}
\end{array}\right) 
\end{eqnarray}
where

\begin{eqnarray*} \fl
\begin{array}{ll}

K_1=&D_x^9+\alpha_{1}D_x^7+D_x^7\alpha_{1}+\alpha_{2}D_x^5+D_x^5\alpha_{2}+\alpha_{3}D_x^3+D_x^3\alpha_{3}+\alpha_{4}D_x+D_x\alpha_{4}\\&+\beta_{01}\beta_{1}D_x^{-1}\beta_{2}+\beta_{01}\beta_{2}D_x^{-1}\beta_{1}+\beta_{02}\beta_{3}D_x^{-1}\beta_{4}+\beta_{02}\beta_{4}D_x^{-1}\beta_{3}\\\\

K_2=&\frac{11}{18}D_x^9+D_x^8\alpha_{5}+D_x^7\alpha_{6}+D_x^6\alpha_{7}+D_x^5\alpha_{8}+D_x^4\alpha_{9}+D_x^3\alpha_{10}+D_x^2\alpha_{11}\\&+D_x\alpha_{12}+\alpha_{13}+\beta_{01}\beta_{1}D_x^{-1}\beta_{5}+\beta_{01}\beta_{2}D_x^{-1}\beta_{6}+\beta_{02}\beta_{3}D_x^{-1}\beta_{7}+\beta_{02}\beta_{4}D_x^{-1}\beta_{8}\\\\

K_4=&\frac{7}{18}D_x^9+\alpha_{14} D_x^7+D_x^7\alpha_{14} +\alpha_{15} D_x^5+D_x^5\alpha_{15} +\alpha_{16} D_x^3+D_x^3\alpha_{16} +\alpha_{17} D_x \\&+D_x\alpha_{17} +\beta_{01}\beta_{6}D_x^{-1}\beta_{5}+\beta_{01}\beta_{5}D_x^{-1}\beta_{6}+\beta_{02}\beta_{8}D_x^{-1}\beta_{7}+\beta_{02}\beta_{7}D_x^{-1}\beta_{8}

\end{array}.
\end{eqnarray*}

The explicit form of $\alpha_i$ and $\beta_i$ are given in the appendix.
 Almost all of the known integrable models possess linear Lax pairs. However in the literature, there is no systematic way of finding whether a given evolution equation possesses a Lax representation and how one can construct the operators $\mathcal{L} $, and $\mathcal{M}$. For the system (\ref{nm}) we shall consider the case where
our candidate Lax pair is a differential operator

\begin{eqnarray}\fl\begin{array}{ll}
\mathcal{M} =& \alpha_{0}\mathcal{D}_x^{i}+\alpha_{1}\mathcal{D}_x^{i-1}+ ...+\alpha_{i-1}\mathcal{D}_x+\alpha_{i}\\\\ 
\mathcal{L} =& \beta_{0}\mathcal{D}_x^{j}+\beta_{1}\mathcal{D}_x^{j-1}+...+\beta_{j-1}\mathcal{D}_x+\beta_{j}.
\end{array}\end {eqnarray}

Here $\alpha = \alpha(u, v, u_x, v_x, ...)$ and $\beta = \beta(u, v, u_x, v_x, ...)$ are the dependent variables and $i,j= 1,2,3,...$ . Now the major problem is to find appropriate analytic functions $\alpha_i$ and $\beta_j$ that satisfy the
integrability condition (\ref{lax}). From the order of the system (\ref{nm}) it is easy to see that we must assign $i=3$
\begin{eqnarray}\fl
\mathcal{M} =\alpha_{0}\mathcal{D}_x^{3}+\alpha_{1}\mathcal{D}_x^{2}+\alpha_{2}\mathcal{D}_x+\alpha_{3}.
\end {eqnarray}
By assuming that $\alpha_{0}$ and $\beta_{0}$ are nonzero constants, We carry out the calculations For cases $j=1,2,3,\dots$ according to (\ref{lax}). After lengthy calculations, the integrability condition (\ref{lax}) can lead to the following Lax pair:

\begin{eqnarray}\fl\begin{array}{ll}
\mathcal{M}=& -8/3D_x^3 - (4u+4v) D_x^2 - ( 4u_{x}+4v_{x}+2u^{2}+4uv\\&+2v^{2}) D_x + \frac{1}{6}(-260u_{xx}-152v_{xx}-30u_{x}u \\&-48u_{x}v+6v_{x}u +24v_{x}v-9u^{2}v-9uv^{2}),\\
\mathcal{L}_{_{j=1}}=& 2D_x +u+v, \\

\mathcal{L}_{_{j=2}}=& D_x^2 +(u+v) D_x + \frac{1}{4}(2u_{x}+2v_{x}+u^{2}+2uv+v^{2}), \\

\mathcal{L}_{_{j=3}}=& 2/3 D_x^3 + (u+v) D_x^2+ (u_{x}+v_{x}+\frac{1}{2}u^{2}+uv\\&+\frac{1}{2}v^{2}) D_x + \frac{1}{12}(4u_{xx}+4v_{xx}+6u_{x}u+6u_{x}v\\&+6v_{x}u+6v_{x}v+u^{3}+3u^{2}v+3uv^{2}+v^{3})\\
\qquad \vdots &

\end{array}\end {eqnarray}

Using the previous result we will state a conjecture about the existence a hierarchy of infinitely many Lax operators of the new system.

\emph{{\bf Conjecture 2.3.}} Consider the differential operators

\begin{eqnarray*}\fl\begin{array}{ll}
\mathcal{M}=& -8/3D_x^3 - (4u+4v) D_x^2 - ( 4u_{x}+4v_{x}+2u^{2}+4uv+2v^{2}) D_x\\& + \frac{1}{6}(-260u_{xx}-152v_{xx}-30u_{x}u -48u_{x}v+6v_{x}u +24v_{x}v\\&-9u^{2}v-9uv^{2}),
\end{array}\end {eqnarray*}
\begin{eqnarray}\fl
\mathcal{L}_j=\sum_{n=0}^{j}\beta_{_n}D_x^{^{j-n}}, ~ j=1,2,3, \dots \infty , 
\end {eqnarray}

where 
\begin{eqnarray*}\fl
\beta_n= \frac{j-n+1}{2n}(2\frac{d}{dx}+u+v)\beta_{n-1} ,~ \beta_0=\frac{2}{j} .\hspace{4.5cm} 
\end {eqnarray*}

Then it can be further proved that the operator equation

\begin{eqnarray}\fl\begin{array}{ll}
{\mathcal{L}_{j}}_{t}-[\mathcal{M},\mathcal{L}_{j}] =\mathcal{O},
\end{array}\end {eqnarray}

where $\mathcal{O}$ is the zero operator, is equivalent to the system (\ref{nm}) in the sense that both sides of the equation turn out to be operators of multiplication by a function. So the system (\ref{nm}) admits a hierarchy of infinitely many Lax operators that satisfy the condition $[\mathcal{L}_n,\mathcal{L}_m]=0$.

\section*{Appendix}

The explicit form of $\alpha_i$ and $\beta_i$ are :

$\fl\\
\alpha_{1}=(348u_{1}+168v_{1}-61u^{2}-28v^{2})/1296\\
\alpha_{2}=(-233280u_{3}-190080v_{3}+36360u_{2}u+5040v_{2}u+63360v_{2}v+59256u_{1}^{2}+33264u_{1}v_{1}-2364u_{1}u^{2}-5544u_{1}v^{2}+68904v_{1}^{2}-588v_{1}u^{2}-1680v_{1}uv-1848v_{1}v^{2}+439u^{4}+98u^{2}v^{2}+154v^{4})/233280\\\\
\alpha_{3}=(17418240u_{5}+13219200v_{5}-3434400u_{4}u-596160v_{4}u-4406400v_{4}v-14624064u_{3}u_{1}-2032128u_{3}v_{1}+142128u_{3}u^{2}+338688u_{3}v^{2}-3328128v_{3}u_{1}-18605376v_{3}v_{1}+171936v_{3}u^{2}+198720v_{3}uv+163296v_{3}v^{2}-12095568u_{2}^{2}-5225472u_{2}v_{2}+862272u_{2}u_{1}u-18144u_{2}v_{1}u+1741824u_{2}v_{1}v-26856u_{2}u^{3}+3024u_{2}uv^{2}-14362272v_{2}^{2}+356832v_{2}u_{1}u+1109376v_{2}u_{1}v+662688v_{2}v_{1}u+1088640v_{2}v_{1}v+1512v_{2}u^{3}-57312v_{2}u^{2}v-11088v_{2}uv^{2}-54432v_{2}v^{3}+67392u_{1}^{3}+181440u_{1}^{2}v_{1}-335772u_{1}^{2}u^{2}-30240u_{1}^{2}v^{2}+1215216u_{1}v_{1}^{2}-20160u_{1}v_{1}u^{2}-118944u_{1}v_{1}uv-35280u_{1}v_{1}v^{2}+1380u_{1}u^{4}+3360u_{1}u^{2}v^{2}+2940u_{1}v^{4}+338688v_{1}^{3}-65124v_{1}^{2}u^{2}-22176v_{1}^{2}uv-187488v_{1}^{2}v^{2}-1344v_{1}u^{4}-504v_{1}u^{3}v+2604v_{1}u^{2}v^{2}+3696v_{1}uv^{3}+1008v_{1}v^{4}-335u^{6}+224u^{4}v^{2}-217u^{2}v^{4}-56v^{6})/8398080\\\\
\alpha_{4}=(-1917561600u_{7}-1427673600v_{7}+312595200u_{6}u+18662400v_{6}u+475891200v_{6}v+2293608960u_{5}u_{1}+215550720u_{5}v_{1}-15318720u_{5}u^{2}-35925120u_{5}v^{2}+411505920v_{5}u_{1}+2959856640v_{5}v_{1}-12363840v_{5}u^{2}-6220800v_{5}uv-17418240v_{5}v^{2}+6056881920u_{4}u_{2}+878999040u_{4}v_{2}-176670720u_{4}u_{1}u+3265920u_{4}v_{1}u-292999680u_{4}v_{1}v+10018080u_{4}u^{3}-544320u_{4}uv^{2}+1200925440v_{4}u_{2}+7548007680v_{4}v_{2}-118661760v_{4}u_{1}u-137168640v_{4}u_{1}v-36547200v_{4}v_{1}u-171383040v_{4}v_{1}v+2721600v_{4}u^{3}+4121280v_{4}u^{2}v+907200v_{4}uv^{2}+5806080v_{4}v^{3}+4166380800u_{3}^{2}+1511654400u_{3}v_{3}-352408320u_{3}u_{2}u-24416640u_{3}v_{2}u-503884800u_{3}v_{2}v-279780480u_{3}u_{1}^{2}-18195840u_{3}u_{1}v_{1}+108889920u_{3}u_{1}u^{2}+3032640u_{3}u_{1}v^{2}-513293760u_{3}v_{1}^{2}+2747520u_{3}v_{1}u^{2}+8138880u_{3}v_{1}uv+3136320u_{3}v_{1}v^{2}-110160u_{3}u^{4}-457920u_{3}u^{2}v^{2}-261360u_{3}v^{4}+5094835200v_{3}^{2}-85380480v_{3}u_{2}u-400308480v_{3}u_{2}v-88335360v_{3}v_{2}u-360495360v_{3}v_{2}v-127370880v_{3}u_{1}^{2}-595330560v_{3}u_{1}v_{1}+9292320v_{3}u_{1}u^{2}+39553920v_{3}u_{1}uv+7776000v_{3}u_{1}v^{2}-369360000v_{3}v_{1}^{2}+17405280v_{3}v_{1}u^{2}+10523520v_{3}v_{1}uv+70761600v_{3}v_{1}v^{2}-395280v_{3}u^{4}-907200v_{3}u^{3}v-153360v_{3}u^{2}v^{2}-302400v_{3}uv^{3}-168480v_{3}v^{4}-322237440u_{2}^{2}u_{1}-45567360u_{2}^{2}v_{1}+110211840u_{2}^{2}u^{2}+7594560u_{2}^{2}v^{2}-183358080u_{2}v_{2}u_{1}-1261111680u_{2}v_{2}v_{1}+7698240u_{2}v_{2}u^{2}+28460160u_{2}v_{2}uv+10031040u_{2}v_{2}v^{2}+417467520u_{2}u_{1}^{2}u+9175680u_{2}u_{1}v_{1}u+61119360u_{2}u_{1}v_{1}v-902880u_{2}u_{1}u^{3}-1529280u_{2}u_{1}uv^{2}+28071360u_{2}v_{1}^{2}u+20062080u_{2}v_{1}^{2}v-237600u_{2}v_{1}u^{3}-2566080u_{2}v_{1}u^{2}v+129600u_{2}v_{1}uv^{2}-3343680u_{2}v_{1}v^{3}-111600u_{2}u^{5}+39600u_{2}u^{3}v^{2}-10800u_{2}uv^{4}-450230400v_{2}^{2}u_{1}-528145920v_{2}^{2}v_{1}+15577920v_{2}^{2}u^{2}+8709120v_{2}^{2}uv+57075840v_{2}^{2}v^{2}+22109760v_{2}u_{1}^{2}u+42456960v_{2}u_{1}^{2}v+124960320v_{2}u_{1}v_{1}u+41368320v_{2}u_{1}v_{1}v+360720v_{2}u_{1}u^{3}-3097440v_{2}u_{1}u^{2}v-1049760v_{2}u_{1}uv^{2}-2592000v_{2}u_{1}v^{3}+15240960v_{2}v_{1}^{2}u+222549120v_{2}v_{1}^{2}v-2857680v_{2}v_{1}u^{3}-2449440v_{2}v_{1}u^{2}v-4173120v_{2}v_{1}uv^{2}-2229120v_{2}v_{1}v^{3}-32760v_{2}u^{5}+131760v_{2}u^{4}v+22680v_{2}u^{3}v^{2}+51120v_{2}u^{2}v^{3}+30240v_{2}uv^{4}+56160v_{2}v^{5}+80193888u_{1}^{4}+1378944u_{1}^{3}v_{1}-692064u_{1}^{3}u^{2}-229824u_{1}^{3}v^{2}+46878912u_{1}^{2}v_{1}^{2}+930096u_{1}^{2}v_{1}u^{2}-7369920u_{1}^{2}v_{1}uv-1473984u_{1}^{2}v_{1}v^{2}+759492u_{1}^{2}u^{4}-155016u_{1}^{2}u^{2}v^{2}+122832u_{1}^{2}v^{4}+16609536u_{1}v_{1}^{3}-2794608u_{1}v_{1}^{2}u^{2}-2099520u_{1}v_{1}^{2}uv-7423488u_{1}v_{1}^{2}v^{2}+22608u_{1}v_{1}u^{4}-120240u_{1}v_{1}u^{3}v-100944u_{1}v_{1}u^{2}v^{2}+349920u_{1}v_{1}uv^{3}+88128u_{1}v_{1}v^{4}-1476u_{1}u^{6}-3768u_{1}u^{4}v^{2}+8412u_{1}u^{2}v^{4}-4896u_{1}v^{6}+39408768v_{1}^{4}-339552v_{1}^{3}u^{2}-3265920v_{1}^{3}uv-2270592v_{1}^{3}v^{2}+161388v_{1}^{2}u^{4}+45360v_{1}^{2}u^{3}v+424656v_{1}^{2}u^{2}v^{2}+120960v_{1}^{2}uv^{3}+325728v_{1}^{2}v^{4}+7908v_{1}u^{6}+10920v_{1}u^{5}v-9876v_{1}u^{4}v^{2}-7560v_{1}u^{3}v^{3}-2736v_{1}u^{2}v^{4}-10080v_{1}uv^{5}-1152v_{1}v^{6}+663u^{8}-1318u^{6}v^{2}+823u^{4}v^{4}+152u^{2}v^{6}+48v^{8})/1511654400\\
\alpha_{5}=(-11u-22v)/108\\
\alpha_{6}=(300u_{1}+114v_{1}-19u^{2}+22uv-19v^{2})/648\\
\alpha_{7}=(-4176u_{2}-1656v_{2}+1380u_{1}u-468u_{1}v-114v_{1}u+324v_{1}v+19u^{3}+38u^{2}v+19uv^{2}+38v^{3})/3888\\
\alpha_{8}=(234360u_{3}+51840v_{3}-106380u_{2}u+27720u_{2}v+8280v_{2}u-7560v_{2}v-99468u_{1}^{2}+24408u_{1}v_{1}-6948u_{1}u^{2}-11520u_{1}uv-4068u_{1}v^{2}-13932v_{1}^{2}+324v_{1}u^{2}-1620v_{1}uv-4356v_{1}v^{2}+93u^{4}-190u^{3}v-54u^{2}v^{2}-190uv^{3}+93v^{4})/116640\\\\
\alpha_{9}=(-594000u_{4}+321840u_{3}u-100800u_{3}v-17280v_{3}u-15120v_{3}v+918360u_{2}u_{1}-109800u_{2}v_{1}+27120u_{2}u^{2}+47880u_{2}uv+18300u_{2}v^{2}-126720v_{2}u_{1}-28080v_{2}v_{1}-3660v_{2}u^{2}+2520v_{2}uv+9720v_{2}v^{2}+49236u_{1}^{2}u+43272u_{1}^{2}v-2496u_{1}v_{1}u+28248u_{1}v_{1}v-3064u_{1}u^{3}+3492u_{1}u^{2}v+416u_{1}uv^{2}+2332u_{1}v^{3}+4644v_{1}^{2}u+12168v_{1}^{2}v-108v_{1}u^{3}+1004v_{1}u^{2}v+1452v_{1}uv^{2}-816v_{1}v^{3}-31u^{5}-62u^{4}v+18u^{3}v^{2}+36u^{2}v^{3}-31uv^{4}-62v^{5})/233280\\\\
\alpha_{10}=(8981280u_{5}-2138400v_{5}-5702400u_{4}u+1749600u_{4}v+440640v_{4}v-21394368u_{3}u_{1}+2603664u_{3}v_{1}-624024u_{3}u^{2}-1069200u_{3}uv-433944u_{3}v^{2}+2843424v_{3}u_{1}+3334608v_{3}v_{1}+133272v_{3}u^{2}+45360v_{3}uv+10152v_{3}v^{2}-16000416u_{2}^{2}+4802976u_{2}v_{2}-2949480u_{2}u_{1}u-3090528u_{2}u_{1}v+45144u_{2}v_{1}u-1194048u_{2}v_{1}v+95256u_{2}u^{3}-99864u_{2}u^{2}v-7524u_{2}uv^{2}-67824u_{2}v^{3}+2586816v_{2}^{2}+435456v_{2}u_{1}u-497664v_{2}u_{1}v+84240v_{2}v_{1}u-99792v_{2}v_{1}v+10980v_{2}u^{3}-26352v_{2}u^{2}v-29160v_{2}uv^{2}+12168v_{2}v^{3}-831168u_{1}^{3}+67608u_{1}^{2}v_{1}+281916u_{1}^{2}u^{2}-169704u_{1}^{2}uv-11268u_{1}^{2}v^{2}-842832u_{1}v_{1}^{2}+3168u_{1}v_{1}u^{2}-137952u_{1}v_{1}uv-185040u_{1}v_{1}v^{2}+7716u_{1}u^{4}+13920u_{1}u^{3}v-528u_{1}u^{2}v^{2}-1200u_{1}uv^{3}+2916u_{1}v^{4}-152280v_{1}^{3}-42372v_{1}^{2}u^{2}-36504v_{1}^{2}uv+42876v_{1}^{2}v^{2}-342v_{1}u^{4}-3012v_{1}u^{3}v-6708v_{1}u^{2}v^{2}+2448v_{1}uv^{3}+5634v_{1}v^{4}-41u^{6}+186u^{5}v+57u^{4}v^{2}-108u^{3}v^{3}+57u^{2}v^{4}+186uv^{5}-41v^{6})/4199040\\\\
\alpha_{11}=(-28926720u_{6}+13530240v_{6}+22122720u_{5}u-7464960u_{5}v+2138400v_{5}u-1866240v_{5}v+99042912u_{4}u_{1}-11749536u_{4}v_{1}+2956176u_{4}u^{2}+4989600u_{4}uv+1958256u_{4}v^{2}-8833536v_{4}u_{1}-23654592v_{4}v_{1}-736128v_{4}u^{2}-440640v_{4}uv-697248v_{4}v^{2}+191056320u_{3}u_{2}-29160000u_{3}v_{2}+17385840u_{3}u_{1}u+17830368u_{3}u_{1}v-252720u_{3}v_{1}u+6954336u_{3}v_{1}v-551880u_{3}u^{3}+648864u_{3}u^{2}v+42120u_{3}uv^{2}+460944u_{3}v^{3}-29276640v_{3}u_{2}-48600000v_{3}v_{2}-4612464v_{3}u_{1}u+230688v_{3}u_{1}v-3334608v_{3}v_{1}u-2428704v_{3}v_{1}v-133272v_{3}u^{3}+87264v_{3}u^{2}v-10152v_{3}uv^{2}+11664v_{3}v^{3}+11036736u_{2}^{2}u+13457664u_{2}^{2}v-3701376u_{2}v_{2}u+5295456u_{2}v_{2}v+18505584u_{2}u_{1}^{2}-370656u_{2}u_{1}v_{1}-4818744u_{2}u_{1}u^{2}+2664144u_{2}u_{1}uv+61776u_{2}u_{1}v^{2}+8906112u_{2}v_{1}^{2}+5832u_{2}v_{1}u^{2}+1186704u_{2}v_{1}uv+1772064u_{2}v_{1}v^{2}-57024u_{2}u^{4}-106992u_{2}u^{3}v-972u_{2}u^{2}v^{2}+7848u_{2}uv^{3}-23688u_{2}v^{4}-2586816v_{2}^{2}u-1845504v_{2}^{2}v-3551040v_{2}u_{1}^{2}+6941376v_{2}u_{1}v_{1}-384696v_{2}u_{1}u^{2}+926640v_{2}u_{1}uv+1219968v_{2}u_{1}v^{2}+2076192v_{2}v_{1}^{2}+704808v_{2}v_{1}u^{2}+99792v_{2}v_{1}uv-116640v_{2}v_{1}v^{2}+4860v_{2}u^{4}+26352v_{2}u^{3}v+57924v_{2}u^{2}v^{2}-12168v_{2}uv^{3}-29736v_{2}v^{4}-3016224u_{1}^{3}u+644112u_{1}^{3}v+159624u_{1}^{2}v_{1}u+1091664u_{1}^{2}v_{1}v-203868u_{1}^{2}u^{3}-313272u_{1}^{2}u^{2}v-26604u_{1}^{2}uv^{2}+15336u_{1}^{2}v^{3}+1491264u_{1}v_{1}^{2}u+1365552u_{1}v_{1}^{2}v+24768u_{1}v_{1}u^{3}+133560u_{1}v_{1}u^{2}v+219024u_{1}v_{1}uv^{2}-48528u_{1}v_{1}v^{3}+4236u_{1}u^{5}-9852u_{1}u^{4}v-4128u_{1}u^{3}v^{2}-888u_{1}u^{2}v^{3}-1284u_{1}uv^{4}-4716u_{1}v^{5}+152280v_{1}^{3}u-195696v_{1}^{3}v+42372v_{1}^{2}u^{3}+59040v_{1}^{2}u^{2}v-42876v_{1}^{2}uv^{2}-77976v_{1}^{2}v^{3}+342v_{1}u^{5}-936v_{1}u^{4}v+6708v_{1}u^{3}v^{2}-372v_{1}u^{2}v^{3}-5634v_{1}uv^{4}+324v_{1}v^{5}+41u^{7}+82u^{6}v-57u^{5}v^{2}-114u^{4}v^{3}-57u^{3}v^{4}-114u^{2}v^{5}+41uv^{6}+82v^{7})/25194240\\\\
\alpha_{12}=(132969600u_{7}-87480000v_{7}-136468800u_{6}u+32659200u_{6}v-33825600v_{6}u+1166400v_{6}v-658782720u_{5}u_{1}+69284160u_{5}v_{1}-20781360u_{5}u^{2}-35769600u_{5}uv-11547360u_{5}v^{2}-7698240v_{5}u_{1}+174726720v_{5}v_{1}+4704480v_{5}u^{2}+4665600v_{5}uv+9370080v_{5}v^{2}-1543030560u_{4}u_{2}+184524480u_{4}v_{2}-146266560u_{4}u_{1}u-152915040u_{4}u_{1}v+2138400u_{4}v_{1}u-44245440u_{4}v_{1}v+4827600u_{4}u^{3}-5047920u_{4}u^{2}v-356400u_{4}uv^{2}-2877120u_{4}v^{3}+136702080v_{4}u_{2}+434367360v_{4}v_{2}+44867520v_{4}u_{1}u+10108800v_{4}u_{1}v+59136480v_{4}v_{1}u+46694880v_{4}v_{1}v+1840320v_{4}u^{3}-259200v_{4}u^{2}v+1743120v_{4}uv^{2}+162000v_{4}v^{3}-996105600u_{3}^{2}+246110400u_{3}v_{3}-228983760u_{3}u_{2}u-284096160u_{3}u_{2}v+36586080u_{3}v_{2}u-44673120u_{3}v_{2}v-191134080u_{3}u_{1}^{2}+3265920u_{3}u_{1}v_{1}+52151040u_{3}u_{1}u^{2}-27501120u_{3}u_{1}uv-544320u_{3}u_{1}v^{2}-73988640u_{3}v_{1}^{2}+42120u_{3}v_{1}u^{2}-11638080u_{3}v_{1}uv-15137280u_{3}v_{1}v^{2}+625320u_{3}u^{4}+1154520u_{3}u^{3}v-7020u_{3}u^{2}v^{2}-92880u_{3}uv^{3}+223560u_{3}v^{4}+289850400v_{3}^{2}+67554000v_{3}u_{2}u-6842880v_{3}u_{2}v+121500000v_{3}v_{2}u+95372640v_{3}v_{2}v+61236000v_{3}u_{1}^{2}+27254880v_{3}u_{1}v_{1}+10014840v_{3}u_{1}u^{2}-5799600v_{3}u_{1}uv-5346000v_{3}u_{1}v^{2}+2702160v_{3}v_{1}^{2}-6065280v_{3}v_{1}u^{2}+6071760v_{3}v_{1}uv-408240v_{3}v_{1}v^{2}-63180v_{3}u^{4}-218160v_{3}u^{3}v-470880v_{3}u^{2}v^{2}-29160v_{3}uv^{3}-24300v_{3}v^{4}-234543600u_{2}^{2}u_{1}-4607280u_{2}^{2}v_{1}+38465280u_{2}^{2}u^{2}-15221520u_{2}^{2}uv+767880u_{2}^{2}v^{2}+90201600u_{2}v_{2}u_{1}-112324320u_{2}v_{2}v_{1}+5559840u_{2}v_{2}u^{2}-13880160u_{2}v_{2}uv-16971120u_{2}v_{2}v^{2}+138464640u_{2}u_{1}^{2}u-23580720u_{2}u_{1}^{2}v-6285600u_{2}u_{1}v_{1}u-31680720u_{2}u_{1}v_{1}v+5867100u_{2}u_{1}u^{3}+9880920u_{2}u_{1}u^{2}v+1047600u_{2}u_{1}uv^{2}+268920u_{2}u_{1}v^{3}-22057920u_{2}v_{1}^{2}u-19316880u_{2}v_{1}^{2}v-404460u_{2}v_{1}u^{3}-1802520u_{2}v_{1}u^{2}v-3032640u_{2}v_{1}uv^{2}+781920u_{2}v_{1}v^{3}-66870u_{2}u^{5}+137340u_{2}u^{4}v+67410u_{2}u^{3}v^{2}-8460u_{2}u^{2}v^{3}+12780u_{2}uv^{4}+47700u_{2}v^{5}+22744800v_{2}^{2}u_{1}+4898880v_{2}^{2}v_{1}-4571640v_{2}^{2}u^{2}+4613760v_{2}^{2}uv-466560v_{2}^{2}v^{2}+7853760v_{2}u_{1}^{2}u-13199760v_{2}u_{1}^{2}v-43655760v_{2}u_{1}v_{1}u-20191680v_{2}u_{1}v_{1}v-642060v_{2}u_{1}u^{3}-2148120v_{2}u_{1}u^{2}v-3254040v_{2}u_{1}uv^{2}+436320v_{2}u_{1}v^{3}-5190480v_{2}v_{1}^{2}u-1632960v_{2}v_{1}^{2}v-1762020v_{2}v_{1}u^{3}-1433160v_{2}v_{1}u^{2}v+291600v_{2}v_{1}uv^{2}+71280v_{2}v_{1}v^{3}-12150v_{2}u^{5}+7020v_{2}u^{4}v-144810v_{2}u^{3}v^{2}+5940v_{2}u^{2}v^{3}+74340v_{2}uv^{4}-6480v_{2}v^{5}+20124288u_{1}^{4}-2042496u_{1}^{3}v_{1}+5120496u_{1}^{3}u^{2}+5682960u_{1}^{3}uv+340416u_{1}^{3}v^{2}-19288368u_{1}^{2}v_{1}^{2}-1077624u_{1}^{2}v_{1}u^{2}-3256200u_{1}^{2}v_{1}uv-2778624u_{1}^{2}v_{1}v^{2}-310788u_{1}^{2}u^{4}+428220u_{1}^{2}u^{3}v+179604u_{1}^{2}u^{2}v^{2}+106380u_{1}^{2}uv^{3}+31212u_{1}^{2}v^{4}-5247504u_{1}v_{1}^{3}-3181248u_{1}v_{1}^{2}u^{2}-3348000u_{1}v_{1}^{2}uv+1392552u_{1}v_{1}^{2}v^{2}-39132u_{1}v_{1}u^{4}+131220u_{1}v_{1}u^{3}v-449064u_{1}v_{1}u^{2}v^{2}+31320u_{1}v_{1}uv^{3}+203868u_{1}v_{1}v^{4}-7266u_{1}u^{6}-13800u_{1}u^{5}v+6522u_{1}u^{4}v^{2}+13800u_{1}u^{3}v^{3}+4362u_{1}u^{2}v^{4}+3000u_{1}uv^{5}-1926u_{1}v^{6}-579312v_{1}^{4}-56592v_{1}^{3}u^{2}+489240v_{1}^{3}uv+230688v_{1}^{3}v^{2}+22788v_{1}^{2}u^{4}-147600v_{1}^{2}u^{3}v+8856v_{1}^{2}u^{2}v^{2}+194940v_{1}^{2}uv^{3}-18252v_{1}^{2}v^{4}+348v_{1}u^{6}+2340v_{1}u^{5}v+4104v_{1}u^{4}v^{2}+930v_{1}u^{3}v^{3}+2904v_{1}u^{2}v^{4}-810v_{1}uv^{5}-2052v_{1}v^{6}+18u^{8}-205u^{7}v-58u^{6}v^{2}+285u^{5}v^{3}+48u^{4}v^{4}+285u^{3}v^{5}-58u^{2}v^{6}-205uv^{7}+18v^{8})/377913600\\\\
\alpha_{13}=(-52488000u_{8}+41990400v_{8}+97977600u_{7}u-34992000u_{7}v+43740000v_{7}u-10497600v_{7}v+412905600u_{6}u_{1}-33242400u_{6}v_{1}+16038000u_{6}u^{2}+29160000u_{6}uv+5540400u_{6}v^{2}+73483200v_{6}u_{1}-94478400v_{6}v_{1}-2332800v_{6}u^{2}-583200v_{6}uv-1749600v_{6}v^{2}+1066206240u_{5}u_{2}-65085120u_{5}v_{2}+112557600u_{5}u_{1}u+92728800u_{5}u_{1}v-1399680u_{5}v_{1}u+4199040u_{5}v_{1}v-4772520u_{5}u^{3}+5637600u_{5}u^{2}v+233280u_{5}uv^{2}+2916000u_{5}v^{3}+31142880v_{5}u_{2}-263139840v_{5}v_{2}-39074400v_{5}u_{1}u-13996800v_{5}u_{1}v-87363360v_{5}v_{1}u-21345120v_{5}v_{1}v-2352240v_{5}u^{3}-4685040v_{5}uv^{2}+680400v_{5}v^{3}+1679616000u_{4}u_{3}-83980800u_{4}v_{3}+217883520u_{4}u_{2}u+231996960u_{4}u_{2}v-22628160u_{4}v_{2}u+7348320u_{4}v_{2}v+170352720u_{4}u_{1}^{2}-4082400u_{4}u_{1}v_{1}-54344520u_{4}u_{1}u^{2}+33475680u_{4}u_{1}uv+680400u_{4}u_{1}v^{2}+22219920u_{4}v_{1}^{2}-155520u_{4}v_{1}u^{2}+7076160u_{4}v_{1}uv+8806320u_{4}v_{1}v^{2}-729000u_{4}u^{4}-1364040u_{4}u^{3}v+25920u_{4}u^{2}v^{2}+77760u_{4}uv^{3}-160380u_{4}v^{4}-41990400v_{4}u_{3}-430401600v_{4}v_{3}-75349440v_{4}u_{2}u-8281440v_{4}u_{2}v-217183680v_{4}v_{2}u-48055680v_{4}v_{2}v-64851840v_{4}u_{1}^{2}-130986720v_{4}u_{1}v_{1}-15513120v_{4}u_{1}u^{2}+3479760v_{4}u_{1}uv-2080080v_{4}u_{1}v^{2}-25894080v_{4}v_{1}^{2}+3868560v_{4}v_{1}u^{2}-23347440v_{4}v_{1}uv+5598720v_{4}v_{1}v^{2}+58320v_{4}u^{4}+129600v_{4}u^{3}v+262440v_{4}u^{2}v^{2}-81000v_{4}uv^{3}-168480v_{4}v^{4}+135010800u_{3}^{2}u+143817120u_{3}^{2}v-67068000u_{3}v_{3}u-3849120u_{3}v_{3}v+512574480u_{3}u_{2}u_{1}+10905840u_{3}u_{2}v_{1}-97180560u_{3}u_{2}u^{2}+55151280u_{3}u_{2}uv-1817640u_{3}u_{2}v^{2}-86780160u_{3}v_{2}u_{1}+22278240u_{3}v_{2}v_{1}-5239080u_{3}v_{2}u^{2}+13355280u_{3}v_{2}uv+9234000u_{3}v_{2}v^{2}-172257840u_{3}u_{1}^{2}u+37888560u_{3}u_{1}^{2}v+7931520u_{3}u_{1}v_{1}u+22492080u_{3}u_{1}v_{1}v-7414740u_{3}u_{1}u^{3}-12117600u_{3}u_{1}u^{2}v-1321920u_{3}u_{1}uv^{2}+1072440u_{3}u_{1}v^{3}+21889440u_{3}v_{1}^{2}u+9622800u_{3}v_{1}^{2}v+568620u_{3}v_{1}u^{3}+1856520u_{3}v_{1}u^{2}v+3155760u_{3}v_{1}uv^{2}-129600u_{3}v_{1}v^{3}+93690u_{3}u^{5}-213300u_{3}u^{4}v-94770u_{3}u^{3}v^{2}-18360u_{3}u^{2}v^{3}-12960u_{3}uv^{4}-80460u_{3}v^{5}-144925200v_{3}^{2}u-30093120v_{3}^{2}v-169069680v_{3}u_{2}u_{1}-61527600v_{3}u_{2}v_{1}-13967640v_{3}u_{2}u^{2}+10944720v_{3}u_{2}uv+2634120v_{3}u_{2}v^{2}-270021600v_{3}v_{2}u_{1}-110458080v_{3}v_{2}v_{1}+7873200v_{3}v_{2}u^{2}-47686320v_{3}v_{2}uv+7484400v_{3}v_{2}v^{2}-22812840v_{3}u_{1}^{2}u+9486720v_{3}u_{1}^{2}v+51146640v_{3}u_{1}v_{1}u-7367760v_{3}u_{1}v_{1}v+1292760v_{3}u_{1}u^{3}+2284200v_{3}u_{1}u^{2}v+3340440v_{3}u_{1}uv^{2}+288360v_{3}u_{1}v^{3}-1351080v_{3}v_{1}^{2}u+22044960v_{3}v_{1}^{2}v+3032640v_{3}v_{1}u^{3}+903960v_{3}v_{1}u^{2}v+204120v_{3}v_{1}uv^{2}-395280v_{3}v_{1}v^{3}+31590v_{3}u^{5}+1620v_{3}u^{4}v+235440v_{3}u^{3}v^{2}+16740v_{3}u^{2}v^{3}+12150v_{3}uv^{4}-17280v_{3}v^{5}+98910720u_{2}^{3}-56570400u_{2}^{2}v_{2}-243835920u_{2}^{2}u_{1}u+47394720u_{2}^{2}u_{1}v+8009280u_{2}^{2}v_{1}u+18623520u_{2}^{2}v_{1}v-4834080u_{2}^{2}u^{3}-8822520u_{2}^{2}u^{2}v-1334880u_{2}^{2}uv^{2}+38880u_{2}^{2}v^{3}-51438240u_{2}v_{2}^{2}-17982000u_{2}v_{2}u_{1}u+30501360u_{2}v_{2}u_{1}v+75038400u_{2}v_{2}v_{1}u+13160880u_{2}v_{2}v_{1}v+1372140u_{2}v_{2}u^{3}+2818800u_{2}v_{2}u^{2}v+4775760u_{2}v_{2}uv^{2}-288360u_{2}v_{2}v^{3}-133634448u_{2}u_{1}^{3}+11106072u_{2}u_{1}^{2}v_{1}-24477552u_{2}u_{1}^{2}u^{2}-28032480u_{2}u_{1}^{2}uv-1851012u_{2}u_{1}^{2}v^{2}+53355024u_{2}u_{1}v_{1}^{2}+4308552u_{2}u_{1}v_{1}u^{2}+8754480u_{2}u_{1}v_{1}uv+9618912u_{2}u_{1}v_{1}v^{2}+1215324u_{2}u_{1}u^{4}-1649700u_{2}u_{1}u^{3}v-718092u_{2}u_{1}u^{2}v^{2}-460080u_{2}u_{1}uv^{3}-83376u_{2}u_{1}v^{4}+6082776u_{2}v_{1}^{3}+4433292u_{2}v_{1}^{2}u^{2}+4691520u_{2}v_{1}^{2}uv-1427868u_{2}v_{1}^{2}v^{2}+65988u_{2}v_{1}u^{4}-301320u_{2}v_{1}u^{3}v+686556u_{2}v_{1}u^{2}v^{2}+28080u_{2}v_{1}uv^{3}-323082u_{2}v_{1}v^{4}+12744u_{2}u^{6}+25470u_{2}u^{5}v-10998u_{2}u^{4}v^{2}-26010u_{2}u^{3}v^{3}-6183u_{2}u^{2}v^{4}-3780u_{2}uv^{5}+2709u_{2}v^{6}-30093120v_{2}^{3}+38257920v_{2}^{2}u_{1}u-5287680v_{2}^{2}u_{1}v-2449440v_{2}^{2}v_{1}u+31609440v_{2}^{2}v_{1}v+2285820v_{2}^{2}u^{3}+712800v_{2}^{2}u^{2}v+233280v_{2}^{2}uv^{2}-136080v_{2}^{2}v^{3}-6053616v_{2}u_{1}^{3}+61791984v_{2}u_{1}^{2}v_{1}+3526416v_{2}u_{1}^{2}u^{2}+4976640v_{2}u_{1}^{2}uv+4553496v_{2}u_{1}^{2}v^{2}+15050448v_{2}u_{1}v_{1}^{2}+15017724v_{2}u_{1}v_{1}u^{2}+10044000v_{2}u_{1}v_{1}uv+24624v_{2}u_{1}v_{1}v^{2}+189648v_{2}u_{1}u^{4}-205740v_{2}u_{1}u^{3}v+1044846v_{2}u_{1}u^{2}v^{2}+7560v_{2}u_{1}uv^{3}-304452v_{2}u_{1}v^{4}+19397232v_{2}v_{1}^{3}-43416v_{2}v_{1}^{2}u^{2}+816480v_{2}v_{1}^{2}uv+1265544v_{2}v_{1}^{2}v^{2}-63504v_{2}v_{1}u^{4}+716580v_{2}v_{1}u^{3}v+74952v_{2}v_{1}u^{2}v^{2}-35640v_{2}v_{1}uv^{3}-175284v_{2}v_{1}v^{4}-612v_{2}u^{6}-3510v_{2}u^{5}v-6156v_{2}u^{4}v^{2}-2970v_{2}u^{3}v^{3}-6786v_{2}u^{2}v^{4}+3240v_{2}uv^{5}+6858v_{2}v^{6}-6838992u_{1}^{4}u-3075408u_{1}^{4}v+2109240u_{1}^{3}v_{1}u+2145528u_{1}^{3}v_{1}v+1436940u_{1}^{3}u^{3}-1266516u_{1}^{3}u^{2}v-351540u_{1}^{3}uv^{2}-21276u_{1}^{3}v^{3}+7125408u_{1}^{2}v_{1}^{2}u+4263840u_{1}^{2}v_{1}^{2}v+181764u_{1}^{2}v_{1}u^{3}-843048u_{1}^{2}v_{1}u^{2}v+1035504u_{1}^{2}v_{1}uv^{2}+96552u_{1}^{2}v_{1}v^{3}+60138u_{1}^{2}u^{5}+103788u_{1}^{2}u^{4}v-30294u_{1}^{2}u^{3}v^{2}-55404u_{1}^{2}u^{2}v^{3}-13302u_{1}^{2}uv^{4}-22212u_{1}^{2}v^{5}+211248u_{1}v_{1}^{3}u-938952u_{1}v_{1}^{3}v-462024u_{1}v_{1}^{2}u^{3}+1062720u_{1}v_{1}^{2}u^{2}v+30456u_{1}v_{1}^{2}uv^{2}-664740u_{1}v_{1}^{2}v^{3}-10836u_{1}v_{1}u^{5}-49734u_{1}v_{1}u^{4}v-64692u_{1}v_{1}u^{3}v^{2}-5958u_{1}v_{1}u^{2}v^{3}-29016u_{1}v_{1}uv^{4}-6714u_{1}v_{1}v^{5}-738u_{1}u^{7}+2961u_{1}u^{6}v+1806u_{1}u^{5}v^{2}-2247u_{1}u^{4}v^{3}-864u_{1}u^{3}v^{4}-1797u_{1}u^{2}v^{5}+492u_{1}uv^{6}+1311u_{1}v^{7}+289656v_{1}^{4}u+587088v_{1}^{4}v+28296v_{1}^{3}u^{3}+44064v_{1}^{3}u^{2}v-115344v_{1}^{3}uv^{2}-198936v_{1}^{3}v^{3}-11394v_{1}^{2}u^{5}-7020v_{1}^{2}u^{4}v-4428v_{1}^{2}u^{3}v^{2}-6660v_{1}^{2}u^{2}v^{3}+9126v_{1}^{2}uv^{4}+15660v_{1}^{2}v^{5}-174v_{1}u^{7}-144v_{1}u^{6}v-2052v_{1}u^{5}v^{2}+288v_{1}u^{4}v^{3}-1452v_{1}u^{3}v^{4}-762v_{1}u^{2}v^{5}+1026v_{1}uv^{6}+306v_{1}v^{7}-9u^{9}-18u^{8}v+29u^{7}v^{2}+58u^{6}v^{3}-24u^{5}v^{4}-48u^{4}v^{5}+29u^{3}v^{6}+58u^{2}v^{7}-9uv^{8}-18v^{9})/1133740800\\\\
\alpha_{14} =(-36u_{1}-196v_{1}-3u^{2}-14v^{2})/432\\\\
\alpha_{15} =(136080u_{3}+373680v_{3}+22680u_{2}u+6480u_{2}v+11520v_{2}v+31464u_{1}^{2}+40896u_{1}v_{1}+1464u_{1}u^{2}+1080u_{1}uv+1824u_{1}v^{2}+94536v_{1}^{2}+3408v_{1}u^{2}+4248v_{1}v^{2}+61u^{4}+152u^{2}v^{2}+346v^{4})/233280\\\\
\alpha_{16} =(-10108800u_{5}-26982720v_{5}-1684800u_{4}u-933120u_{4}v-1114560v_{4}v-8392896u_{3}u_{1}-5964192u_{3}v_{1}-137808u_{3}u^{2}-155520u_{3}uv-327888u_{3}v^{2}-2477952v_{3}u_{1}-13807584v_{3}v_{1}-206496v_{3}u^{2}-256176v_{3}v^{2}-6838992u_{2}^{2}-7242048u_{2}v_{2}-870480u_{2}u_{1}u-677376u_{2}u_{1}v-994032u_{2}v_{1}u-720576u_{2}v_{1}v-22968u_{2}u^{3}-17568u_{2}u^{2}v-54648u_{2}uv^{2}+4032u_{2}v^{3}-17519328v_{2}^{2}-1207008v_{2}u_{1}u+24192v_{2}u_{1}v-1625184v_{2}v_{1}v+2016v_{2}u^{2}v-11664v_{2}v^{3}-301536u_{1}^{3}-1389744u_{1}^{2}v_{1}-79020u_{1}^{2}u^{2}-35136u_{1}^{2}uv-94104u_{1}^{2}v^{2}-802224u_{1}v_{1}^{2}-65952u_{1}v_{1}u^{2}-120096u_{1}v_{1}uv+8208u_{1}v_{1}v^{2}-540u_{1}u^{4}-2928u_{1}u^{3}v-6576u_{1}u^{2}v^{2}+672u_{1}uv^{3}+2004u_{1}v^{4}-84240v_{1}^{3}-66852v_{1}^{2}u^{2}-325080v_{1}^{2}v^{2}-2748v_{1}u^{4}+684v_{1}u^{2}v^{2}-2628v_{1}v^{4}-15u^{6}-274u^{4}v^{2}+167u^{2}v^{4}-294v^{6})/8398080\\\\
\alpha_{17} =(1105747200u_{7}+2953324800v_{7}+184291200u_{6}u+60652800u_{6}v+97977600v_{6}v+1289571840u_{5}u_{1}+820212480u_{5}v_{1}+15318720u_{5}u^{2}+10108800u_{5}uv+22705920u_{5}v^{2}+265006080v_{5}u_{1}+1715074560v_{5}v_{1}+22083840v_{5}u^{2}+27527040v_{5}v^{2}+3472606080u_{4}u_{2}+2228290560u_{4}v_{2}+148677120u_{4}u_{1}u+73872000u_{4}u_{1}v+136702080u_{4}v_{1}u+236234880u_{4}v_{1}v+2553120u_{4}u^{3}+1944000u_{4}u^{2}v+3784320u_{4}uv^{2}+2462400u_{4}v^{3}+1281173760v_{4}u_{2}+6873361920v_{4}v_{2}+213528960v_{4}u_{1}u-4354560v_{4}u_{1}v+330791040v_{4}v_{1}v-362880v_{4}u^{2}v+5469120v_{4}v^{3}+2394619200u_{3}^{2}+2447107200u_{3}v_{3}+301942080u_{3}u_{2}u+186779520u_{3}u_{2}v+371381760u_{3}v_{2}u+188490240u_{3}v_{2}v+319515840u_{3}u_{1}^{2}+702794880u_{3}u_{1}v_{1}+32063040u_{3}u_{1}u^{2}+18169920u_{3}u_{1}uv+23664960u_{3}u_{1}v^{2}+355752000u_{3}v_{1}^{2}+12998880u_{3}v_{1}u^{2}+39372480u_{3}v_{1}uv+4276800u_{3}v_{1}v^{2}+90720u_{3}u^{4}+324000u_{3}u^{3}v+710640u_{3}u^{2}v^{2}+410400u_{3}uv^{3}+771120u_{3}v^{4}+5441256000v_{3}^{2}+407851200v_{3}u_{2}u+60030720v_{3}u_{2}v+650384640v_{3}v_{2}v+446731200v_{3}u_{1}^{2}+139501440v_{3}u_{1}v_{1}+6480000v_{3}u_{1}u^{2}+10005120v_{3}u_{1}uv-648000v_{3}u_{1}v^{2}+510727680v_{3}v_{1}^{2}+11625120v_{3}v_{1}u^{2}+84291840v_{3}v_{1}v^{2}+270000v_{3}u^{4}-54000v_{3}u^{2}v^{2}+237600v_{3}v^{4}+475657920u_{2}^{2}u_{1}+575190720u_{2}^{2}v_{1}+25453440u_{2}^{2}u^{2}+14281920u_{2}^{2}uv+23600160u_{2}^{2}v^{2}+1334361600u_{2}v_{2}u_{1}+639342720u_{2}v_{2}v_{1}+18351360u_{2}v_{2}u^{2}+31415040u_{2}v_{2}uv-1736640u_{2}v_{2}v^{2}+101723040u_{2}u_{1}^{2}u+30611520u_{2}u_{1}^{2}v+81025920u_{2}u_{1}v_{1}u+155442240u_{2}u_{1}v_{1}v+1185840u_{2}u_{1}u^{3}+4129920u_{2}u_{1}u^{2}v+5503680u_{2}u_{1}uv^{2}+1749600u_{2}u_{1}v^{3}+59292000u_{2}v_{1}^{2}u+2255040u_{2}v_{1}^{2}v+2166480u_{2}v_{1}u^{3}+3110400u_{2}v_{1}u^{2}v+712800u_{2}v_{1}uv^{2}-812160u_{2}v_{1}v^{3}+15120u_{2}u^{5}+32400u_{2}u^{4}v+118440u_{2}u^{3}v^{2}+43200u_{2}u^{2}v^{3}+128520u_{2}uv^{4}-39600u_{2}v^{5}+216328320v_{2}^{2}u_{1}+553340160v_{2}^{2}v_{1}+18027360v_{2}^{2}u^{2}+99766080v_{2}^{2}v^{2}+36702720v_{2}u_{1}^{2}u+29548800v_{2}u_{1}^{2}v+106557120v_{2}u_{1}v_{1}u-5495040v_{2}u_{1}v_{1}v+3058560v_{2}u_{1}u^{3}-311040v_{2}u_{1}u^{2}v-289440v_{2}u_{1}uv^{2}+207360v_{2}u_{1}v^{3}+373636800v_{2}v_{1}^{2}v-457920v_{2}v_{1}u^{2}v+1918080v_{2}v_{1}v^{3}-12960v_{2}u^{4}v+17280v_{2}u^{2}v^{3}-88560v_{2}v^{5}+17751312u_{1}^{4}+29439936u_{1}^{3}v_{1}+1268784u_{1}^{3}u^{2}+3157920u_{1}^{3}uv+1884384u_{1}^{3}v^{2}+75502368u_{1}^{2}v_{1}^{2}+7612704u_{1}^{2}v_{1}u^{2}+6220800u_{1}^{2}v_{1}uv+1011744u_{1}^{2}v_{1}v^{2}+94068u_{1}^{2}u^{4}+129600u_{1}^{2}u^{3}v+574416u_{1}^{2}u^{2}v^{2}+86400u_{1}^{2}uv^{3}+233208u_{1}^{2}v^{4}-979776u_{1}v_{1}^{3}+2701728u_{1}v_{1}^{2}u^{2}+375840u_{1}v_{1}^{2}uv-1894752u_{1}v_{1}^{2}v^{2}+71712u_{1}v_{1}u^{4}+518400u_{1}v_{1}u^{3}v+49824u_{1}v_{1}u^{2}v^{2}-135360u_{1}v_{1}uv^{3}-57168u_{1}v_{1}v^{4}+576u_{1}u^{6}+5400u_{1}u^{5}v+9648u_{1}u^{4}v^{2}+7200u_{1}u^{3}v^{3}+17448u_{1}u^{2}v^{4}-6600u_{1}uv^{5}-14424u_{1}v^{6}+67919472v_{1}^{4}-81648v_{1}^{3}u^{2}+1246752v_{1}^{3}v^{2}+112572v_{1}^{2}u^{4}-157896v_{1}^{2}u^{2}v^{2}+792072v_{1}^{2}v^{4}+1992v_{1}u^{6}+2076v_{1}u^{4}v^{2}-4764v_{1}u^{2}v^{4}+2952v_{1}v^{6}+12u^{8}+268u^{6}v^{2}+727u^{4}v^{4}-1202u^{2}v^{6}+627v^{8})/1511654400\\\\
\beta_{1}=u\\
\beta_{2}=-209952000u_{8}-125971200v_{8}+20995200v_{7}u+41990400v_{7}v-96577920u_{6}u_{1}-46189440u_{6}v_{1}+18195840u_{6}u^{2}+7698240u_{6}v^{2}+58786560v_{6}u_{1}+264539520v_{6}v_{1}+4898880v_{6}u^{2}-6998400v_{6}uv+4898880v_{6}v^{2}-289733760u_{5}u_{2}-138568320u_{5}v_{2}+109175040u_{5}u_{1}u+46189440u_{5}v_{1}v+71383680v_{5}u_{2}+751628160v_{5}v_{2}+2099520v_{5}u_{1}u-19595520v_{5}u_{1}v-37091520v_{5}v_{1}u+53187840v_{5}v_{1}v-816480v_{5}u^{3}-1632960v_{5}u^{2}v-816480v_{5}uv^{2}-1632960v_{5}v^{3}-482889600u_{4}u_{3}-125971200u_{4}v_{3}+235846080u_{4}u_{2}u-7698240u_{4}v_{2}u+41990400u_{4}v_{2}v+204353280u_{4}u_{1}^{2}-46889280u_{4}u_{1}v_{1}+3149280u_{4}u_{1}u^{2}+7814880u_{4}u_{1}v^{2}+33592320u_{4}v_{1}^{2}+583200u_{4}v_{1}u^{2}+2566080u_{4}v_{1}uv+2799360u_{4}v_{1}v^{2}-641520u_{4}u^{4}-97200u_{4}u^{2}v^{2}-233280u_{4}v^{4}-20995200v_{4}u_{3}+1259712000v_{4}v_{3}-5598720v_{4}u_{2}u-23794560v_{4}u_{2}v-88179840v_{4}v_{2}u+113374080v_{4}v_{2}v-20295360v_{4}u_{1}^{2}-109175040v_{4}u_{1}v_{1}-3149280v_{4}u_{1}u^{2}-699840v_{4}u_{1}uv+1866240v_{4}u_{1}v^{2}+114773760v_{4}v_{1}^{2}-9097920v_{4}v_{1}u^{2}-7231680v_{4}v_{1}uv-21228480v_{4}v_{1}v^{2}-116640v_{4}u^{4}+272160v_{4}u^{3}v+155520v_{4}u^{2}v^{2}+272160v_{4}uv^{3}-116640v_{4}v^{4}+144866880u_{3}^{2}u-15396480u_{3}v_{3}u+6998400u_{3}v_{3}v+743230080u_{3}u_{2}u_{1}-93778560u_{3}u_{2}v_{1}+6298560u_{3}u_{2}u^{2}+15629760u_{3}u_{2}v^{2}-109175040u_{3}v_{2}u_{1}-12597120u_{3}v_{2}v_{1}+1166400u_{3}v_{2}u^{2}+5132160u_{3}v_{2}uv+5598720u_{3}v_{2}v^{2}+12597120u_{3}u_{1}^{2}u+2332800u_{3}u_{1}v_{1}u+36391680u_{3}u_{1}v_{1}v-5132160u_{3}u_{1}u^{3}-388800u_{3}u_{1}uv^{2}+5132160u_{3}v_{1}^{2}u+11197440u_{3}v_{1}^{2}v-388800u_{3}v_{1}u^{2}v-1866240u_{3}v_{1}v^{3}-60886080v_{3}^{2}u+69984000v_{3}^{2}v-104976000v_{3}u_{2}u_{1}-123171840v_{3}u_{2}v_{1}-2566080v_{3}u_{2}u^{2}+1866240v_{3}u_{2}uv+4665600v_{3}u_{2}v^{2}-195955200v_{3}v_{2}u_{1}+460494720v_{3}v_{2}v_{1}-16329600v_{3}v_{2}u^{2}-11664000v_{3}v_{2}uv-36158400v_{3}v_{2}v^{2}-3265920v_{3}u_{1}^{2}u+6765120v_{3}u_{1}^{2}v-5598720v_{3}u_{1}v_{1}u+3732480v_{3}u_{1}v_{1}v-116640v_{3}u_{1}u^{3}+1049760v_{3}u_{1}u^{2}v+466560v_{3}u_{1}uv^{2}-622080v_{3}u_{1}v^{3}-11897280v_{3}v_{1}^{2}u-72783360v_{3}v_{1}^{2}v+1244160v_{3}v_{1}u^{3}+311040v_{3}v_{1}u^{2}v+2721600v_{3}v_{1}uv^{2}-1632960v_{3}v_{1}v^{3}+19440v_{3}u^{5}+38880v_{3}u^{4}v-25920v_{3}u^{3}v^{2}-51840v_{3}u^{2}v^{3}+19440v_{3}uv^{4}+38880v_{3}v^{5}+179625600u_{2}^{3}-81881280u_{2}^{2}v_{2}+18895680u_{2}^{2}u_{1}u+1749600u_{2}^{2}v_{1}u+27293760u_{2}^{2}v_{1}v-3849120u_{2}^{2}u^{3}-291600u_{2}^{2}uv^{2}-82581120u_{2}v_{2}^{2}+2099520u_{2}v_{2}u_{1}u+34992000u_{2}v_{2}u_{1}v+2799360u_{2}v_{2}v_{1}u+16796160u_{2}v_{2}v_{1}v-116640u_{2}v_{2}u^{3}+855360u_{2}v_{2}u^{2}v+466560u_{2}v_{2}uv^{2}-1555200u_{2}v_{2}v^{3}+17216064u_{2}u_{1}^{3}-279936u_{2}u_{1}^{2}v_{1}-15544224u_{2}u_{1}^{2}u^{2}+46656u_{2}u_{1}^{2}v^{2}+33032448u_{2}u_{1}v_{1}^{2}+769824u_{2}u_{1}v_{1}u^{2}-699840u_{2}u_{1}v_{1}uv+653184u_{2}u_{1}v_{1}v^{2}-34992u_{2}u_{1}u^{4}-128304u_{2}u_{1}u^{2}v^{2}-54432u_{2}u_{1}v^{4}+8957952u_{2}v_{1}^{3}+1275264u_{2}v_{1}^{2}u^{2}+933120u_{2}v_{1}^{2}uv-2612736u_{2}v_{1}^{2}v^{2}+66096u_{2}v_{1}u^{4}+38880u_{2}v_{1}u^{3}v-139968u_{2}v_{1}u^{2}v^{2}-155520u_{2}v_{1}uv^{3}-31104u_{2}v_{1}v^{4}+11448u_{2}u^{6}-11016u_{2}u^{4}v^{2}+11664u_{2}u^{2}v^{4}+1728u_{2}v^{6}+123871680v_{2}^{3}-4898880v_{2}^{2}u_{1}u-20645280v_{2}^{2}v_{1}u-100077120v_{2}^{2}v_{1}v+738720v_{2}^{2}u^{3}+1652400v_{2}^{2}uv^{2}-1166400v_{2}^{2}v^{3}-559872v_{2}u_{1}^{3}+16936128v_{2}u_{1}^{2}v_{1}+209952v_{2}u_{1}^{2}u^{2}+1088640v_{2}u_{1}^{2}uv+559872v_{2}u_{1}^{2}v^{2}+10077696v_{2}u_{1}v_{1}^{2}+3989088v_{2}u_{1}v_{1}u^{2}+2799360v_{2}u_{1}v_{1}uv-1492992v_{2}u_{1}v_{1}v^{2}+66096v_{2}u_{1}u^{4}+38880v_{2}u_{1}u^{3}v-139968v_{2}u_{1}u^{2}v^{2}-155520v_{2}u_{1}uv^{3}-31104v_{2}u_{1}v^{4}-59066496v_{2}v_{1}^{3}+839808v_{2}v_{1}^{2}u^{2}+6687360v_{2}v_{1}^{2}uv-6858432v_{2}v_{1}^{2}v^{2}+132192v_{2}v_{1}u^{4}-124416v_{2}v_{1}u^{2}v^{2}+194400v_{2}v_{1}uv^{3}+482112v_{2}v_{1}v^{4}+1296v_{2}u^{6}-6480v_{2}u^{5}v-2592v_{2}u^{4}v^{2}+8640v_{2}u^{3}v^{3}-2592v_{2}u^{2}v^{4}-6480v_{2}uv^{5}+1296v_{2}v^{6}-3922992u_{1}^{4}u+513216u_{1}^{3}v_{1}u+186624u_{1}^{3}v_{1}v-46656u_{1}^{3}u^{3}-85536u_{1}^{3}uv^{2}+1508544u_{1}^{2}v_{1}^{2}u+1119744u_{1}^{2}v_{1}^{2}v+132192u_{1}^{2}v_{1}u^{3}-69984u_{1}^{2}v_{1}u^{2}v-139968u_{1}^{2}v_{1}uv^{2}-186624u_{1}^{2}v_{1}v^{3}+34344u_{1}^{2}u^{5}-22032u_{1}^{2}u^{3}v^{2}+11664u_{1}^{2}uv^{4}+933120u_{1}v_{1}^{3}u+373248u_{1}v_{1}^{3}v+38880u_{1}v_{1}^{2}u^{3}-279936u_{1}v_{1}^{2}u^{2}v-466560u_{1}v_{1}^{2}uv^{2}-124416u_{1}v_{1}^{2}v^{3}-22032u_{1}v_{1}u^{4}v+46656u_{1}v_{1}u^{2}v^{3}+10368u_{1}v_{1}v^{5}+789264v_{1}^{4}u-2239488v_{1}^{4}v-46656v_{1}^{3}u^{3}+31104v_{1}^{3}u^{2}v+186624v_{1}^{3}uv^{2}+575424v_{1}^{3}v^{3}-7776v_{1}^{2}u^{5}-5184v_{1}^{2}u^{4}v-2592v_{1}^{2}u^{3}v^{2}-10368v_{1}^{2}u^{2}v^{3}-23976v_{1}^{2}uv^{4}+7776v_{1}^{2}v^{5}-216v_{1}u^{7}-432v_{1}u^{6}v+432v_{1}u^{5}v^{2}+864v_{1}u^{4}v^{3}+432v_{1}u^{3}v^{4}+864v_{1}u^{2}v^{5}-216v_{1}uv^{6}-432v_{1}v^{7}-11u^{9}+36u^{7}v^{2}-36u^{5}v^{4}-24u^{3}v^{6}+9uv^{8}\\\\
\beta_{3}=144u_{2}+108v_{2}-18v_{1}u-36v_{1}v-2u^{3}+3uv^{2}\\
\beta_{4}=2566080u_{6}+1632960v_{6}-272160v_{5}u-544320v_{5}v+816480u_{4}u_{1}+272160u_{4}v_{1}-181440u_{4}u^{2}-45360u_{4}v^{2}-544320v_{4}u_{1}-2449440v_{4}v_{1}-45360v_{4}u^{2}+90720v_{4}uv-45360v_{4}v^{2}+1632960u_{3}u_{2}+544320u_{3}v_{2}-725760u_{3}u_{1}u-181440u_{3}v_{1}v-272160v_{3}u_{2}-4626720v_{3}v_{2}+45360v_{3}u_{1}u+181440v_{3}u_{1}v+317520v_{3}v_{1}u-362880v_{3}v_{1}v+7560v_{3}u^{3}+15120v_{3}u^{2}v+7560v_{3}uv^{2}+15120v_{3}v^{3}-544320u_{2}^{2}u+45360u_{2}v_{2}u+90720u_{2}v_{2}v-852768u_{2}u_{1}^{2}+353808u_{2}u_{1}v_{1}-13608u_{2}u_{1}u^{2}-58968u_{2}u_{1}v^{2}+127008u_{2}v_{1}^{2}+10584u_{2}v_{1}u^{2}-15120u_{2}v_{1}uv-12096u_{2}v_{1}v^{2}+4788u_{2}u^{4}-1764u_{2}u^{2}v^{2}+1008u_{2}v^{4}+226800v_{2}^{2}u-272160v_{2}^{2}v+199584v_{2}u_{1}^{2}+616896v_{2}u_{1}v_{1}+10584v_{2}u_{1}u^{2}-15120v_{2}u_{1}uv-12096v_{2}u_{1}v^{2}-517104v_{2}v_{1}^{2}+51408v_{2}v_{1}u^{2}+45360v_{2}v_{1}uv+127008v_{2}v_{1}v^{2}+756v_{2}u^{4}-2520v_{2}u^{3}v-1008v_{2}u^{2}v^{2}-2520v_{2}uv^{3}+756v_{2}v^{4}-9072u_{1}^{3}u+10584u_{1}^{2}v_{1}u-66528u_{1}^{2}v_{1}v+9576u_{1}^{2}u^{3}-1764u_{1}^{2}uv^{2}-15120u_{1}v_{1}^{2}u-24192u_{1}v_{1}^{2}v-3528u_{1}v_{1}u^{2}v+4032u_{1}v_{1}v^{3}+13608v_{1}^{3}u+81648v_{1}^{3}v-3024v_{1}^{2}u^{3}-2016v_{1}^{2}u^{2}v-6804v_{1}^{2}uv^{2}+3024v_{1}^{2}v^{3}-126v_{1}u^{5}-252v_{1}u^{4}v+168v_{1}u^{3}v^{2}+336v_{1}u^{2}v^{3}-126v_{1}uv^{4}-252v_{1}v^{5}-8u^{7}+21u^{5}v^{2}-14u^{3}v^{4}+7uv^{6}\\\\
\beta_{5}=-125971200u_{8}-83980800v_{8}-20995200u_{7}u-41990400u_{7}v-88179840u_{6}u_{1}-29393280u_{6}v_{1}+4898880u_{6}u^{2}-6998400u_{6}uv+4898880u_{6}v^{2}+33592320v_{6}u_{1}+172160640v_{6}v_{1}+2799360v_{6}u^{2}+13296960v_{6}v^{2}-159563520u_{5}u_{2}-46189440u_{5}v_{2}+56687040u_{5}u_{1}u-22394880u_{5}u_{1}v-4898880u_{5}v_{1}u+5598720u_{5}v_{1}v+816480u_{5}u^{3}+1632960u_{5}u^{2}v+816480u_{5}uv^{2}+1632960u_{5}v^{3}+100776960v_{5}u_{2}+516481920v_{5}v_{2}+16796160v_{5}u_{1}u+79781760v_{5}v_{1}v-230947200u_{4}u_{3}+130870080u_{4}u_{2}u-30792960u_{4}u_{2}v-7698240u_{4}v_{2}u-5598720u_{4}v_{2}v+116173440u_{4}u_{1}^{2}-35691840u_{4}u_{1}v_{1}+9097920u_{4}u_{1}u^{2}+15629760u_{4}u_{1}uv+5948640u_{4}u_{1}v^{2}-4199040u_{4}v_{1}^{2}-933120u_{4}v_{1}u^{2}+933120u_{4}v_{1}uv+3265920u_{4}v_{1}v^{2}-116640u_{4}u^{4}+272160u_{4}u^{3}v+155520u_{4}u^{2}v^{2}+272160u_{4}uv^{3}-116640u_{4}v^{4}+83980800v_{4}u_{3}+860803200v_{4}v_{3}+13996800v_{4}u_{2}u-11197440v_{4}u_{2}v+187557120v_{4}v_{2}v-2799360v_{4}u_{1}^{2}-60186240v_{4}u_{1}v_{1}-2799360v_{4}u_{1}u^{2}-1866240v_{4}u_{1}uv-3032640v_{4}u_{1}v^{2}+76282560v_{4}v_{1}^{2}-5015520v_{4}v_{1}u^{2}-5598720v_{4}v_{1}v^{2}-116640v_{4}u^{4}-252720v_{4}u^{2}v^{2}-524880v_{4}v^{4}+83980800u_{3}^{2}u-23094720u_{3}^{2}v-22394880u_{3}v_{3}v+445098240u_{3}u_{2}u_{1}-65784960u_{3}u_{2}v_{1}+14463360u_{3}u_{2}u^{2}+27993600u_{3}u_{2}uv+10964160u_{3}u_{2}v^{2}-75582720u_{3}v_{2}u_{1}-96577920u_{3}v_{2}v_{1}-3732480u_{3}v_{2}u^{2}-933120u_{3}v_{2}uv+233280u_{3}v_{2}v^{2}+27060480u_{3}u_{1}^{2}u+23094720u_{3}u_{1}^{2}v-4665600u_{3}u_{1}v_{1}u+14929920u_{3}u_{1}v_{1}v-1749600u_{3}u_{1}u^{3}+2216160u_{3}u_{1}u^{2}v+777600u_{3}u_{1}uv^{2}+1710720u_{3}u_{1}v^{3}-699840u_{3}v_{1}^{2}u+933120u_{3}v_{1}^{2}v-155520u_{3}v_{1}u^{3}+933120u_{3}v_{1}u^{2}v+544320u_{3}v_{1}uv^{2}-233280u_{3}v_{1}v^{3}-19440u_{3}u^{5}-38880u_{3}u^{4}v+25920u_{3}u^{3}v^{2}+51840u_{3}u^{2}v^{3}-19440u_{3}uv^{4}-38880u_{3}v^{5}+121072320v_{3}^{2}v-67184640v_{3}u_{2}u_{1}-142767360v_{3}u_{2}v_{1}-5598720v_{3}u_{2}u^{2}-3732480v_{3}u_{2}uv-6065280v_{3}u_{2}v^{2}-120372480v_{3}v_{2}u_{1}+281335680v_{3}v_{2}v_{1}-10031040v_{3}v_{2}u^{2}-11197440v_{3}v_{2}v^{2}-11197440v_{3}u_{1}^{2}u-3732480v_{3}u_{1}^{2}v-23794560v_{3}u_{1}v_{1}u-12130560v_{3}u_{1}v_{1}v-933120v_{3}u_{1}u^{3}-1010880v_{3}u_{1}uv^{2}-22394880v_{3}v_{1}^{2}v-1010880v_{3}v_{1}u^{2}v-4199040v_{3}v_{1}v^{3}+111274560u_{2}^{3}-59486400u_{2}^{2}v_{2}+35341920u_{2}^{2}u_{1}u+37091520u_{2}^{2}u_{1}v-2799360u_{2}^{2}v_{1}u+13996800u_{2}^{2}v_{1}v-1283040u_{2}^{2}u^{3}+1302480u_{2}^{2}u^{2}v+466560u_{2}^{2}uv^{2}+972000u_{2}^{2}v^{3}-107075520u_{2}v_{2}^{2}-19595520u_{2}v_{2}u_{1}u+8398080u_{2}v_{2}u_{1}v-16096320u_{2}v_{2}v_{1}u-16796160u_{2}v_{2}v_{1}v-622080u_{2}v_{2}u^{3}+933120u_{2}v_{2}u^{2}v+38880u_{2}v_{2}uv^{2}-233280u_{2}v_{2}v^{3}+20435328u_{2}u_{1}^{3}-6158592u_{2}u_{1}^{2}v_{1}-7138368u_{2}u_{1}^{2}u^{2}+4587840u_{2}u_{1}^{2}uv+1026432u_{2}u_{1}^{2}v^{2}+19175616u_{2}u_{1}v_{1}^{2}-559872u_{2}u_{1}v_{1}u^{2}+5598720u_{2}u_{1}v_{1}uv+5738688u_{2}u_{1}v_{1}v^{2}-225504u_{2}u_{1}u^{4}-427680u_{2}u_{1}u^{3}v+93312u_{2}u_{1}u^{2}v^{2}+155520u_{2}u_{1}uv^{3}-89424u_{2}u_{1}v^{4}+2519424u_{2}v_{1}^{3}+1772928u_{2}v_{1}^{2}u^{2}+155520u_{2}v_{1}^{2}uv-559872u_{2}v_{1}^{2}v^{2}+15552u_{2}v_{1}u^{4}+155520u_{2}v_{1}u^{3}v+342144u_{2}v_{1}u^{2}v^{2}-38880u_{2}v_{1}uv^{3}-101088u_{2}v_{1}v^{4}+1296u_{2}u^{6}-6480u_{2}u^{5}v-2592u_{2}u^{4}v^{2}+8640u_{2}u^{3}v^{3}-2592u_{2}u^{2}v^{4}-6480u_{2}uv^{5}+1296u_{2}v^{6}+68351040v_{2}^{3}-17845920v_{2}^{2}u_{1}u-9097920v_{2}^{2}u_{1}v-33592320v_{2}^{2}v_{1}v-758160v_{2}^{2}u^{2}v-3149280v_{2}^{2}v^{3}-6718464v_{2}u_{1}^{3}-419904v_{2}u_{1}^{2}v_{1}-1446336v_{2}u_{1}^{2}u^{2}+1866240v_{2}u_{1}^{2}uv+2091744v_{2}u_{1}^{2}v^{2}+6158592v_{2}u_{1}v_{1}^{2}+2612736v_{2}u_{1}v_{1}u^{2}-2799360v_{2}u_{1}v_{1}uv-419904v_{2}u_{1}v_{1}v^{2}+15552v_{2}u_{1}u^{4}+155520v_{2}u_{1}u^{3}v+342144v_{2}u_{1}u^{2}v^{2}-38880v_{2}u_{1}uv^{3}-101088v_{2}u_{1}v^{4}-32752512v_{2}v_{1}^{3}+513216v_{2}v_{1}^{2}u^{2}-10878624v_{2}v_{1}^{2}v^{2}+108864v_{2}v_{1}u^{4}-34992v_{2}v_{1}u^{2}v^{2}+69984v_{2}v_{1}v^{4}+432v_{2}u^{6}+14256v_{2}u^{4}v^{2}-8424v_{2}u^{2}v^{4}+10152v_{2}v^{6}-2192832u_{1}^{4}u+555984u_{1}^{4}v-93312u_{1}^{3}v_{1}u+1866240u_{1}^{3}v_{1}v-256608u_{1}^{3}u^{3}-365472u_{1}^{3}u^{2}v+15552u_{1}^{3}uv^{2}+62208u_{1}^{3}v^{3}+3545856u_{1}^{2}v_{1}^{2}u+2208384u_{1}^{2}v_{1}^{2}v+62208u_{1}^{2}v_{1}u^{3}+466560u_{1}^{2}v_{1}u^{2}v+684288u_{1}^{2}v_{1}uv^{2}-38880u_{1}^{2}v_{1}v^{3}+7776u_{1}^{2}u^{5}-21384u_{1}^{2}u^{4}v-10368u_{1}^{2}u^{3}v^{2}+2592u_{1}^{2}u^{2}v^{3}-5184u_{1}^{2}uv^{4}-11664u_{1}^{2}v^{5}+419904u_{1}v_{1}^{3}u-279936u_{1}v_{1}^{3}v+295488u_{1}v_{1}^{2}u^{3}+342144u_{1}v_{1}^{2}u^{2}v-93312u_{1}v_{1}^{2}uv^{2}-202176u_{1}v_{1}^{2}v^{3}+2592u_{1}v_{1}u^{5}+57024u_{1}v_{1}u^{3}v^{2}-16848u_{1}v_{1}uv^{4}+216u_{1}u^{7}+432u_{1}u^{6}v-432u_{1}u^{5}v^{2}-864u_{1}u^{4}v^{3}-432u_{1}u^{3}v^{4}-864u_{1}u^{2}v^{5}+216u_{1}uv^{6}+432u_{1}v^{7}-2290032v_{1}^{4}v-23328v_{1}^{3}u^{2}v+93312v_{1}^{3}v^{3}+14256v_{1}^{2}u^{4}v-16848v_{1}^{2}u^{2}v^{3}+30456v_{1}^{2}v^{5}+9u^{8}v-24u^{6}v^{3}-36u^{4}v^{5}+36u^{2}v^{7}-11v^{9}\\\\
\beta_{6}=v\\
\beta_{7}=1632960u_{6}+933120v_{6}+272160u_{5}u+544320u_{5}v+816480u_{4}u_{1}+272160u_{4}v_{1}-45360u_{4}u^{2}+90720u_{4}uv-45360u_{4}v^{2}-1360800v_{4}v_{1}-136080v_{4}v^{2}+816480u_{3}u_{2}+272160u_{3}v_{2}-408240u_{3}u_{1}u+181440u_{3}u_{1}v+45360u_{3}v_{1}u-7560u_{3}u^{3}-15120u_{3}u^{2}v-7560u_{3}uv^{2}-15120u_{3}v^{3}-2721600v_{3}v_{2}-544320v_{3}v_{1}v-317520u_{2}^{2}u+45360u_{2}^{2}v+45360u_{2}v_{2}u-480816u_{2}u_{1}^{2}+208656u_{2}u_{1}v_{1}-57456u_{2}u_{1}u^{2}-105840u_{2}u_{1}uv-34776u_{2}u_{1}v^{2}+27216u_{2}v_{1}^{2}+6048u_{2}v_{1}u^{2}-9072u_{2}v_{1}v^{2}+756u_{2}u^{4}-2520u_{2}u^{3}v-1008u_{2}u^{2}v^{2}-2520u_{2}uv^{3}+756u_{2}v^{4}-408240v_{2}^{2}v+81648v_{2}u_{1}^{2}+54432v_{2}u_{1}v_{1}+6048v_{2}u_{1}u^{2}-9072v_{2}u_{1}v^{2}-308448v_{2}v_{1}^{2}+4536v_{2}v_{1}u^{2}+27216v_{2}v_{1}v^{2}+252v_{2}u^{4}-756v_{2}u^{2}v^{2}+4032v_{2}v^{4}-34776u_{1}^{3}u-18144u_{1}^{3}v+12096u_{1}^{2}v_{1}u+3024u_{1}^{2}u^{3}-5796u_{1}^{2}u^{2}v-2016u_{1}^{2}uv^{2}-4536u_{1}^{2}v^{3}+4536u_{1}v_{1}^{2}u-9072u_{1}v_{1}^{2}v+1008u_{1}v_{1}u^{3}-1512u_{1}v_{1}uv^{2}+126u_{1}u^{5}+252u_{1}u^{4}v-168u_{1}u^{3}v^{2}-336u_{1}u^{2}v^{3}+126u_{1}uv^{4}+252u_{1}v^{5}+18144v_{1}^{3}v-756v_{1}^{2}u^{2}v+8064v_{1}^{2}v^{3}+7u^{6}v-14u^{4}v^{3}+21u^{2}v^{5}-8v^{7}\\
\beta_{8}=108u_{2}+36v_{2}+18u_{1}u+36u_{1}v+3u^{2}v-2v^{3}\\
\beta_{01}=1/13604889600\\
\beta_{02}=(-1)/13604889600.\\
$

Also the recursion operator for this equation can be written in the form $R=\mathrm{J}\circ\mathrm{K}$.

\address{Department of Engineering Physics, Ankara University 06100 Tando\u{g}an, Ankara}
 
\end{document}